\documentstyle[12pt]{article}
\makeatletter
\if@twoside
\else
\fi
\ifcase \@ptsize
\or 
\or 
\fi

\def\@citex[#1]#2{%
\if@filesw \immediate \write \@auxout {\string \citation {#2}}\fi
\@tempcntb\m@ne \let\@h@ld\relax \def\@citea{}%
\@cite{%
 \@for \@citeb:=#2\do {%
 \@ifundefined {b@\@citeb}%
 {\@h@ld\@citea\@tempcntb\m@ne{\bf ?}%
  \@warning {Citation `\@citeb ' on page \thepage \space undefined}}%
 {\@tempcnta\@tempcntb \advance\@tempcnta\@ne%
 \@tempcntb\number\csname b@\@citeb \endcsname \relax%
 \ifnum\@tempcnta=\@tempcntb %
 \ifx\@h@ld\relax%
 \edef \@h@ld{\@citea\csname b@\@citeb\endcsname}%
 \else%
 \edef\@h@ld{\ifmmode{-}\else--\fi\csname b@\@citeb\endcsname}%
 \fi%
 \else
 \@h@ld\@citea\csname b@\@citeb \endcsname%
        \let\@h@ld\relax%
      \fi}%
    \def\@citea{,\penalty\@highpenalty\,}%
  }\@h@ld
}{#1}}
\@addtoreset{equation}{section}
\def\section{\@startsection {section}{1}{\z@}{-3.5ex plus -1ex minus
 -.2ex}{2.3ex plus .2ex}{\large\bf\centering}}
\def\subsection{\@startsection{subsection}{2}{\z@}{-3.25ex plus%
 -1ex minus -.2ex}{1.5ex plus .2ex}{\sc}}
\gdef\@publabel{\hfil}
\gdef\@pubdate{\null}
\gdef\@pubnumber{\null}
\gdef\@author{\null}
\gdef\@title{\null}
\gdef\@abstract{\null}
\long\def\pubdate#1{\gdef\@pubdate{#1}}
\long\def\pubnumber#1{\gdef\@pubnumber{#1}}
\long\def\publabel#1{\gdef\@publabel{#1}}
\long\def\author#1{\gdef\@author{#1}}
\long\def\title#1{\gdef\@title{#1}}
\long\def\abstract#1{\gdef\@abstract{#1}}
\def\titlerelax{
\let\maketitle\relax
\let\settitleparameters\relax
\let\consolidatetitle\relax
\let\inittitlepage\relax
\let\finishtitlepage\relax
\let\titlepagecontents\relax
\let\multithanks\relax
\let\titlebaselines\relax
\let\@makepub\relax
\let\@maketitle\relax
\let\@makeauthor\relax
\let\@makeabstract\relax
\let\@maketitlenote\relax
\let\thanks\relax
\let\titlerelax\relax}
\def\titleclean
{\gdef\@titlenote{}
\gdef\@abstract{}
\gdef\@author{}
\gdef\@title{}
\gdef\@pubdate{}\gdef\@pubnumber{}\gdef\@publabel{}
\gdef\@dpublabel{}
}

\def\@makepub{\vbox to \z@{\hbox to \textwidth{\hfill
\@publabel \hfill
\llap{\parbox[t]{0.33\textwidth}{\raggedleft\@pubnumber}}}%
\vss}}
\def\@maketitle{\vskip 60pt \begin{center}
 {\LARGE \@title \par}
 \end{center}}

\def\@makeauthor{{%
\def\and{\smallskip {\normalsize \rm and\smallskip }}
\def\And{\medskip {\normalsize \rm and\\}\medskip}
\long\def\address##1{{\def\and{\\and\\}\medskip
				{\small \it \\##1\\}
}}
{\centering
 \vskip 3em
 \large \lineskip .75em
 \@author}
 \par}}
\def\@makedate{\vskip 1.5em
 {\raggedright \small \noindent\@pubdate \par}}
\def\@makeabstract{\vskip 1.5em
{\small
\begin{center}
{\bf ABSTRACT\vspace{-.5em}\vspace{0pt}}
\end{center}
\quotation \@abstract \endquotation}}
\def\maketitle{\titlepage
\let\footnotesize\small \setcounter{page}{0}
\@makepub
\vfil
\@maketitle
\@makeauthor
\vfil
\@makeabstract
\@thanks
\vfil
\@makedate
\if@restonecol\twocolumn \else \eject \fi
\titlerelax \titleclean
\setcounter{footnote}{0}
}

\def\bigans{y }

\read-1 to\answ
\ifx\answ\bigans
\read-1 to\bnsw
	\ifx\bnsw\bigans 

\ifcase\@ptsize
 \font\tenmsa=msam10
 \font\sevenmsa=msam7
 \font\fivemsa=msam5
 \font\tenmsb=msbm10
 \font\sevenmsb=msbm7
 \font\fivemsb=msbm5
\or
 \font\tenmsa=msam10 scaled \magstephalf
 \font\sevenmsa=msam8
 \font\fivemsa=msam6
 \font\tenmsb=msbm10 scaled \magstephalf
 \font\sevenmsb=msbm8
 \font\fivemsb=msbm6
\or
 \font\tenmsa=msam10 scaled \magstep1
 \font\sevenmsa=msam8
 \font\fivemsa=msam6
 \font\tenmsb=msbm10 scaled \magstep1
 \font\sevenmsb=msbm8
 \font\fivemsb=msbm6
\fi

\newfam\msafam
\newfam\msbfam
\textfont\msafam=\tenmsa  \scriptfont\msafam=\sevenmsa
  \scriptscriptfont\msafam=\fivemsa
\textfont\msbfam=\tenmsb  \scriptfont\msbfam=\sevenmsb
  \scriptscriptfont\msbfam=\fivemsb

\def\hexnumber@#1{\ifnum#1<10 \number#1\else
 \ifnum#1=10 A\else\ifnum#1=11 B\else\ifnum#1=12 C\else
 \ifnum#1=13 D\else\ifnum#1=14 E\else\ifnum#1=15 F\fi\fi\fi\fi\fi\fi\fi}

\def\msa@{\hexnumber@\msafam}
\def\msb@{\hexnumber@\msbfam}
\def\Bbb{\ifmmode\let\next\Bbb@\else
 \def\next{\errmessage{Use \string\Bbb\space only in math mode}}\fi\next}
\def\Bbb@#1{{\Bbb@@{#1}}}
\def\Bbb@@#1{\fam\msbfam#1}
	\else
		\ifx\cnsw\bigans 

\ifcase\@ptsize
 \font\tenmsx=msxm10
 \font\sevenmsx=msxm7
 \font\fivemsx=msxm5
 \font\tenmsy=msym10
 \font\sevenmsy=msym7
 \font\fivemsy=msym5
\or
 \font\tenmsx=msxm10 scaled \magstephalf
 \font\sevenmsx=msxm8
 \font\fivemsx=msxm6
 \font\tenmsy=msym10 scaled \magstephalf
 \font\sevenmsy=msym8
 \font\fivemsy=msym6
\or
 \font\tenmsx=msxm10 scaled \magstep1
 \font\sevenmsx=msxm8
 \font\fivemsx=msxm6
 \font\tenmsy=msym10 scaled \magstep1
 \font\sevenmsy=msym8
 \font\fivemsy=msym6
\fi

\newfam\msxfam
\newfam\msyfam
\textfont\msxfam=\tenmsx  \scriptfont\msxfam=\sevenmsx
  \scriptscriptfont\msxfam=\fivemsx
\textfont\msyfam=\tenmsy  \scriptfont\msyfam=\sevenmsy
  \scriptscriptfont\msyfam=\fivemsy

\def\hexnumber@#1{\ifnum#1<10 \number#1\else
 \ifnum#1=10 A\else\ifnum#1=11 B\else\ifnum#1=12 C\else
 \ifnum#1=13 D\else\ifnum#1=14 E\else\ifnum#1=15 F\fi\fi\fi\fi\fi\fi\fi}

\def\msx@{\hexnumber@\msxfam}
\def\msy@{\hexnumber@\msyfam}
\def\Bbb{\ifmmode\let\next\Bbb@\else
 \def\next{\errmessage{Use \string\Bbb\space only in math mode}}\fi\next}
\def\Bbb@#1{{\Bbb@@{#1}}}
\def\Bbb@@#1{\fam\msyfam#1}

\else
\def\Bbb#1{{\bf #1}}
		\fi
	\fi
\else
\def\Bbb#1{{\bf #1}}
\fi

\def\cN{{\cal N}}
\def\cO{{\cal O}}
\def\cP{{\cal P}}
\let\a=\alpha
\def\alm{\alpha _-}
\def\alp{\alpha_+}
\def\alm{\alpha_-}

\def\cev#1{\langle #1 \vert}
\def\cdd{{\cdot}}
\def\cO{{\cal O}}
\def\comm#1#2{\bigl [ #1 , #2 \bigl ] }
\def\cont{\nonumber\\*&&\mbox{}}
\def\half{{1\over2}}
\def\half#1{\frac{#1}{2}}
\def\vac{\vec 0}
\def\vec#1{\vert #1 \rangle}

\def\WA{\mathop{\it WA}\nolimits}

\def\en{\end{equation}}
\def\eq{\begin{equation}}
\def\eqq{\begin{eqnarray}}
\def\enn{\end{eqnarray}}

\relax
\@writefile{toc}{\string\contentsline\space {section}%
{\string\numberline\space {1}Introduction}{1}}
\@writefile{toc}{\string\contentsline\space {section}%
{\string\numberline\space {2}Null vectors in the Virasoro theory}{1}}
\@writefile{toc}{\string\contentsline\space {section}%
{\string\numberline\space {3}Null vectors in the $W_3$ algebra}{3}}
\@writefile{toc}{\string\contentsline\space {section}%
{\string\numberline\space {4}Conclusions}{7}}

\makeatother

\begin{document}
\pubnumber{EFI 92--50 \\DAMTP 92--62}
\pubdate{September 1992}
\title{Null vectors of the $W_3$ Algebra}

\author{P.~BOWCOCK$^{1,2}$
\address{
Enrico Fermi Institute,
University of Chicago,
Chicago, IL 60637, U.S.A.}
\And
G.~M.~T.~WATTS$^{3,4}$
\address{
DAMTP, University of Cambridge,
Silver Street, Cambridge, CB3 9EW,
United Kingdom}
}

\footnotetext[1]{
Email: {\tt BOWCOCK@YUKAWA.UCHICAGO.EDU}
}
\footnotetext[2]{
Supported by DOE of USA grant Number DEFG02-90-ER-40560 and NSF grant
PHY900036
}
\footnotetext[3]{
Email: {\tt G.M.T.Watts@DAMTP.CAMBRIDGE.AC.UK}
}
\footnotetext[4]{
Supported by a research fellowship from St John's College, Cambridge,
U.K.
}

\abstract{
We
construct $W_3$ null vectors of a restricted class
explicitly in two different forms.
The method we use is an extension of that of Bauer et al.~in the
Virasoro case. Our results are analogous to
the formulae of Benoit and St.~Aubin for the Virasoro null vectors. We
derive in the Virasoro case some alternative
formulae for the same null vectors involving only the $L_{-1}$ and $L_{-2}$
modes of the Virasoro algebra.
}

\maketitle

\section{Introduction}

A crucial step in the study of rational conformal field theories is to
understand degenerate representations of the Virasoro algebra and its
extensions.
Much of the analysis to date relies on free field representations of
these algebras to calculate the embedding structure of null vectors,
determine the unitary representations and calculate the correlation
functions of the corresponding rational models
\cite{FField}.
These methods only give implicit expressions for the null
vectors. More recently however, explicit and surprisingly elegant
formulae for a subset of the null vectors were found by Benoit and
St-Aubin \cite{BStA1}. These formulae were explained
by Bauer et al.
\cite{BDIZ1} and extended by Kent \cite{Kent5}. These approaches
only utilised information intrinsic to the Virasoro algebra.
It is the aim of this letter to extend the above work to W-algebras.

We start with a brief review of the results of \cite{BDIZ1}
for the Virasoro algebra representations. In this paper the formulae for
null vectors of the type $(1,p)$ and $(q,1)$ are obtained by
considering the operator product expansion (ope) of a null vector in one
representation with the highest weight in another.
Surprisingly, we are able to derive new formulae for the Virasoro null
states involving only the two generating modes $L_{-1}$ and $L_{-2}$
by expanding the operator product about a different point.

We find that to extend these
results to other chiral algebras we need to consider several such
opes simultaneously. We concentrate on the case of $W_3$ for simplicity,
deriving explicit formulae for a corresponding subset of null vectors.

\section{Null vectors in the Virasoro theory}

The generators of infinitesimal conformal transformations satisfy
the Virasoro algebra
\eq
\comm{L_m}{L_n} =
(c/12)m(m^2-1)\delta_{m+n,0} + (m-n)L_{m+n}.
\label{eq.vir}
\en
An irreducible highest weight representation of the  Virasoro algebra
with highest weight state $\vec h$ such that
$L_m\vec h = h\delta_{m,0}\vec h\;,\; m\ge 0$,
is uniquely determined by the values of $c$ and $h$.
Whenever $h,c$ satisfy
\eq
c = c(t) = 13 - 6t - 6/t
\;,\;\;
h = h_{p,q}(t) = (1/4t)( (pt - q)^2 - (t-1)^2 )
\,.
\en
for some $t$ and
for $p,q \in \Bbb Z$
there is a null state at level $pq$ which we denote by
$\cO_{p,q}\vec{h_{p,q}}$,
which is itself a highest weight state.
This state and all states formed by acting on it with modes of $L$ are
orthogonal to all the states in the representation; we call these null
states.
We denote the  primary field corresponding to the states $\vec{h_{p,q}}$
by $\phi_{p,q}$ where
\eq
 \lim_{z \to 0} \phi_{p,q}(z)\vac = \vec{h_{p,q}}
\,.
\en
The field corresponding to a state $L_{-p}\vec\phi$ for any state
$\phi$ we write as
$\hat L_{-p}\phi(z)$, so that a basis for the fields in the theory is
given by the fields $\hat L_{-p_1}\ldots\hat L_{-p_n}\phi(z)$
where $\phi(z)$ runs over the primary fields.

It is natural to require that the three point function of two primary
fields and a null descendant of a third primary field vanish. By
considering this Feigin and Fuchs gave the condition for the coupling
of three primary fields not to vanish \cite{FeFu3}, when one of the
representations has a null state.
Using their formula we find that the three point function
\eq
\cev{h_{q,1}} \hat\cO_{1,2}\phi_{1,2}(z) \vec{h_{q,0}} = 0
\;.
\label{eq.tpf}
\en
In particular,
if we consider the restriction of the operator product expansion of
the null field at level 2 in the representation  $(1,2)$ with the
primary field $\phi_{q,0}$
to the representation $h_{q,1}$, then the coefficient of the highest
weight itself must vanish. So, if the ope does not vanish identically,
the leading term must be a null highest weight state in the $(q,1)$
representation. In ref.~\cite{BDIZ1} Bauer et al.~found that is indeed
the case
and that this provides an explicit form for the null state.

More explicitly, the null state at level 2 in the representation
$h_{1,2}$ is
\eq
\vec{\chi} =  \left [L_{-2} - t L_{-1}^2 \right ] \vec{h_{1,2}}
\,.
\en
Using contour manipulations it is possible to write the ope of a
descendant state of an arbitrary field $\phi$ of conformal weight
$\Delta_\phi$ with a highest weight
state $\vec h$ in two ways,
\eq \hat L_{-p}\phi(z) \vec h
= \left[
\sum_{s=0}^\infty  \pmatrix{ p-2-s \cr s } z^s L_{-p-s}
+ (-1/z)^p  ( h (p-1) + z L_{-1} - z\partial)
\right]
\phi(z)\vec h
\label{eq.bsa}
\en
or alternatively,
\eq
\phi_h(z)  L_{-p}\vec{\phi} =
\left [ L_{-p}
 - z^{-p+1}\partial + (p-1) h z^{-p}
\right ] \phi_h(z) \vec\phi
\,.
\label{eq.bw}
\en

Using (\ref{eq.bsa}) Bauer et al.~re-wrote the operator product
expansion of the field $\chi(z)$
with $\phi_{q,0}$
in terms of the ope of $\phi_{1,2}(z)$ with $\phi_{q,0}$ as
\eq
\chi(z)\vec{h_{q,0}}
= \left[
\sum_{s=0}^\infty   z^s L_{-2-s}
+ (1/z)^2  ( h_{q,0}  + z L_{-1} - z\partial)
- t \partial^2
\right]
\phi(z)\vec{ h_{q,0} }
\label{eq.bsa2}
\,.
\en
Expanding the two opes as
\eq
\phi_{1,2}(z)\vec{h_{q,0}} =
\sum_{n=0}^\infty z^{n-y} f_n
\;,\;
\chi(z)\vec{h_{q,0}} =
\sum_{n=0}^\infty z^{n-y-2} \tilde f_n
\;,\;
\en
where $y= -h_{q,1} + h_{1,2} + h_{q,0}$,
we find from (\ref{eq.bsa2}) that
\eq
\tilde f_n
=
- t n( n-q) f_n
+
\sum_{i > 0} L_{-i} f_{n-i}
\,.
\en
We see that $\tilde f_0=0$ as required, and that setting $\tilde f_i,
O<i<q$ to zero we can recursively solve for $f_i, i<q$.
We obtain an equation for $\tilde f_q$, which yields the
Benoit-St.~Aubin form for the null state at level $q$,
\eq
 \cO_{q,1}\vec{q,1} =
 \tilde f_q =
\sum_{{\{m_1,\ldots,m_r\}}\atop{\sum m_i = q}}
\;
\frac{
 \prod_{i=1}^r  L_{-m_i}
}{
 \prod_{i=1}^{r-1} t ( \sum_{j=1}^i m_j )( \sum_{j=1}^i m_j - q )
}
\vec{q,1}
\,.
\en
Here we note that if instead we use the form (\ref{eq.bw}) and
consider
$\phi_{q,0}(z)\vec\chi$, we obtain
an alternative expression for this null vector in
terms of $L_{-1}, L_{-2}$ only,
\eq
\cO_{q,1}\vec{q,1} =
\sum_{{\{m_1,\ldots,m_r\}\,,\, m_i=1,2 }\atop{ \sum m_i = q}}
\;
\frac{
 \prod_{i=1}^r \alpha_{N_i}^{m_i} \tilde L_{-m_i}
}{
 \prod_{i=1}^{r-1}  t ( \sum_{j=1}^i m_j )( \sum_{j=1}^i m_j - q )
}
\vec{q,1}
\en
where the numerator is ordered from right to left, $N_i=\sum_{j=1}^i m_i$
and we have used
\eqq
\tilde L_{-1} = L_{-1} \;,\;& \tilde L_{-2} = L_{-2} - t L_{-1}^2
\nonumber\\
\alpha_N^1 = 1 + t(1 + q - 2N) \;,\;&
\alpha_N^2 = 1
\enn

\section{Null vectors in the $W_3$ algebra}

\def\al{\alpha}

The $W_3$ algebra is the simplest of the non-linear extended chiral
algebras, and was first written down by Zamolodchikov \cite{Zamo1}.
We give it here
in a normalisation which will be useful later on. The generators are
$L_m, Q_m$ satisfying (\ref{eq.vir}) and
\eqq
\comm{ L_m}{Q_n} &=& (2m-n)Q_{m+n} \nonumber\\
\comm{ Q_m}{Q_n} &=&
\frac{(22 + 5c)}{48}\frac{c}{3\cdot 5!} (m^2-4)(m^2-1)m\delta_{m+n}
\\
&& + \frac{1}{3}(m-n)\Lambda_{m+n} +
\frac{(22 + 5c)}{48}\frac{(m-n)}{30}(2m^2-mn+2n^2-8)L_{m+n}
\,,
\nonumber\enn
where $\vec\Lambda = (L_{-2}L_{-2} - (3/5)L_{-4})\vac$. This
is related to the usual normalisation by
$
Q = \sqrt{ \frac{ 22 + 5c}{48} } W.
$
The representation theory of the $W_3$ algebra can be developed in
analogy with that of the Virasoro algebra. A $W_3$ highest weight
vector $\vec{h,q}$ satisfies
\eq
L_m\vec{h,q}=\delta_{m,0}h\vec{h,q}\;,\;
Q_m\vec{h,q}=\delta_{m,0}q\vec{h,q}\;,\;
m \ge 0\,.
\en
We can parametrise the weights
of a W-highest weight vector as follows \cite{FZam4},
\eq
h=\frac{1}{3}( x^2 + xy + y^2 - 3 a^2)
\;,\; w
 = \frac{1}{27}(x-y)(2x+y)(x+2y)
\,,
\label{eq.weig}
\en
where we define $a,\alpha_\pm$ by
\eq
c= 2 - 24 a^2 \;,\; \alpha_\pm^2 - \alpha_\pm a - 1 = 0
\,.
\en

The determinant formula \cite{WDet} indicates which representations
have null vectors and at what levels they occur. Let
$\lambda^i$ be the two fundamental weights of $A_2$, and form the
vector
$
\mu = x \lambda^1 + y \lambda^2.
$
If
$
\mu\cdd e = r\alp+s\alm
$
for some root $e$ of $A_2$ then there is a null state at level
$rs$.
For generic $c$, a representation has zero, one or two independent
null vectors.  Doubly
degenerate primary fields are given by those representations with two
independent null vectors. Such representations have
\eq x = a \alpha - c/\alpha
\;,\; y = b\alpha - d/\alpha
\,.
\en
for integer $a,b,c,d$, and we denote them by
$
\phi_{ a ,b ; c,d },
$
or simply $(a b; c d)$.

The generalisations of equations (\ref{eq.bsa},\ref{eq.bw}) are
\eqq
{\hat Q}_{-p}\phi(z) \vec{h,q}
&=&
(-1)^{p+3} z^{-p}\left[
\half{(p-1)(p-2)}q - z(p-2){\hat Q}_{-1} - z^2(p-1){\hat Q}_{-2}
\right] \phi(z)\vec{h,q}
\cont
-(-1/z)^p \left[ z Q_{-1} + z^2 Q_{-2} \right]
 \phi(z)\vec{h,q}
\cont+
\sum_{s=0}^{\infty}
\pmatrix{ p-3 - s \cr s}
\left[
z^s Q_{-p-s} \phi(z)\vec{h,q}
\right]
\,,
\label{eq.w}
\enn
and
\eq
\phi_{h,q}(z) Q_{-p}\vec\phi
=
\phi_{h,q}(z)
\left[
z^{-p+2}(p-1)Q_{-2}
-z^{-p+1}(p-2)Q_{-1}
\right]
\vec\phi
\label{eq.walt}
\en
$$
+
\left[
Q_{-p}
+ \half{(p-1)(p-2)}qz^{-p}
+ z^{-p+1}(p-2) Q_{-1}
- z^{-p+2}(p-1) Q_{-2}
\right]
\phi_{h,q}(z)\vec\phi
\,.
$$
If we wish to generalise the construction of null vectors in the
previous section by fusing a single null state with a highest weight
state, then we are still left with the unknown opes
$
\hat Q_{-1}\phi_{h,q}(z)\vec\phi \,,\;
\hat Q_{-2}\phi_{h,q}(z)\vec\phi$.
In the Virasoro case we did indeed get
such opes, but were able to use
$\hat L_{-1}\phi(z) = \partial\phi(z)$.
Here we have no such geometrical interpretation.

In a different context, Bajnok, Palla and Takacs \cite{BPTa1}
considered the Toda theory of which the $W_3$
algebra is a symmetry.
They found a field in classical $A_2$ Toda theory which satisfies a
third order differential equation of motion${}^1$.
\footnotetext[1]{This is the analogue of the
Liouville theory equation in the footnote on p.~357 in
ref.~\cite{BPZ}}
When they attempted to quantise the field they found that
this equation implied the presence of two independent null
vectors at levels one and two, and so this field must be of type
$(2,1;1,1), (1,2;1,1), (1,1;2,1)$ or $(1,1;1,2)$.
Using both the null vectors together with the original equation which
corresponds to a  descendent null vector, they solved for the allowed
fusions of this field with other W-primary fields.  They also found
that the four point functions of this field satisfied a differential
equation, as suggested in ref.~\cite{FZam4}.

This work indicates that the key to understanding the fusion  and
operator product expansions of $\WA_n$ primary fields in more
generality is to consider the null vectors and their descendents
simultaneously.  We found that given  a `completely degenerate'
$\WA_n$ primary field (in the notation of \cite{FLuk})
one can solve not only for the allowed fusions of this field but also
(under some conditions) for the whole ope with another W-primary field.
We intend to discuss these topics in a later work~\cite{BWat3}.
In this letter we shall restrict ourselves to the construction of null
vectors of the $\WA_2$ algebra, in which case the completely
degenerate representations are the doubly degenerate representations
$(ab;cd)$.

We consider the fusion
\eq
(21;11) \,\times\,(01;ss')\;\to\;(11;ss')
\,.
\en
Using (\ref{eq.w}), we can reexpress the ope of some descendant
$\hat {\cal W}\phi_{(21;11)}(z)\vec{01;ss'}$ at level $N$
in terms of the basis opes
\eq
P(m,n)=(\hat Q_{-1})^m(\hat Q_{-2})^n \phi_{21;11}(z)\vec{01;ss'}
\;,
\en
where $m+2n\le N$.

Using the equations (\ref{eq.bsa}) and (\ref{eq.w}), we see that
the opes of all the null states can be related to the subset
\eq
\cN(m,n,i) =
(\hat Q_{-1})^m(\hat Q_{-2})^n \hat\cO_{-i}\phi_{21;11} \vec{01;ss'}
\,,
\label{eq.iope}
\en
where $\cO_{-1}\vec{21;11}$, $\cO_{-2}\vec{21;11}$ are the two
highest weight vectors in the representation $(21;11)$ at level one
and two respectively.
Even the $\cN(m,n,i)$ are not fully independent, being related by further
identities arising from the embedded null vectors at
levels four, five and six.
{}From now on we shall denote the state $\vec{21;11}$ by $\vec\phi$.

\def\yyy{\phi}

If we count the equations arising from the null vectors (\ref{eq.iope}),
and the unknown opes $P(m,n)$, then we see that
at level one there is only one null vector, while
there are two basis opes $P(0,0)$, $P(1,0)$. Up to level two there are
three null vectors to consider and four basis opes. However,
up to level three, there are six null vectors to consider and six
basis vectors, and this balance persists to all higher levels (see
\cite{BWat3}).

The two highest weight null vectors at levels one and two are
\eqq
\cO_{-1}\vec{\yyy} &=&
[
\a Q_{-1}-{5\a^2-3 \over 6}L_{-1}
]
\vec\phi
\label{eq.n12}
\\
\cO_{-2}\vec{\yyy} &=&
[
 L_{-1}^2(7\al^4 - 24\al^2 + 9)
- 12L_{-1}Q_{-1}\al( - 4\al^2 + 3)
\cont+ 24L_{-2}\al^2( - 2\al^4 + 3\al^2 - 1)
+ 36Q_{-1}^2\al^2
- 6Q_{-2}\al(12\al^4 - 13\al^2 + 3)
] \vec\phi
\,.
\nonumber
\enn
The null vectors up to level 6 are then
\eq
\cN=
{ \{
\cO_{-1}\vec{\yyy},
Q_{-1}\cO_{-1}\vec{\yyy},
Q_{-1}^2\cO_{-1}\vec{\yyy}, }
\atop{
Q_{-2}\cO_{-1}\vec{\yyy},
\cO_{-2}\vec{\yyy},
Q_{-1}\cO_{-2}\vec{\yyy}
\} }
\,.
\en
Using these, and the six basis vectors up to level 6,
\eq
\cP= \{
P(0,0),P(0,1),P(1,0),P(2,0),P(3,0),P(1,1)
\}
\,,
\en
we have a matrix operator equation,
\eq
M(z) \cP(z) = \cN(z)
\,,
\label{eq.meq}
\en
where $M(z)$ is a $6\times 6$ matrix whose entries are differential
operators in $\partial/\partial z$ whose coefficients are $z$
dependent polynomials in the lowering modes  $Q_{-p}, L_{-p}, p>0$,
and $\cP(z),\cN(z)$ are the opes of the fields corresponding to the
states $\cP,\cN$ with the state $\vec{01;ss'}$.
We can now impose the requirement that $\cN(z)$ is null, which gives
us a recursion relation for $\cP$.

For calculational purposes, it is easiest to use two null vectors
\eqq
\left(
\a Q_{-1}-{5\a^2-3 \over 6}L_{-1}
\right)
\vec\phi &=& \cO_{-1} \vec\phi
\label{eq.vot}\\
\left(
\a Q_{-2}-L_{-1}^2+ {2\a^2 \over 3}L_{-2}
\right)
\vec\phi &=&
{\left( (78 \a^2 - 54)L_{-1} + 36\a Q_{-1}  \right)\cO_{-1}
- \cO_{-2}
}\over{
36(2\a^2-1)(\a^2-1)
}
\vec\phi
 \nonumber
\enn
to relate $P(m,n)$ to $P(0,0)$, and then use the null vector of ref.
\cite{BPTa1}
\eqq
\vec{\tilde{\phi}}&=&
\left(
\a Q_{-3}-\a^{-2}L_{-1}^3 + L_{-1}L_{-2} + {(\a^2-3)\over
6}L_{-3}
\right)
\vec\phi
\\ &=&
(1/ 72\a(2\a^2-1)(\a^2-1))
\Bigl\{
\left( 30 \a (\a^2-3) L_{-1} - 36\a^2 Q_{-1} \right) \cO_{-2}
\cont +
( -24(2\a^2-1)(\a^2-1)\a^2L_{-2}
+ 12\a(4\a^2 +3)L_{-1}Q_{-1}
+36\a^2Q_{-1}^2
\cont
-6\a(12\a^4-31\a^2+15)Q_{-2}
+(7\a^4 +132\a^2 -99)L_{-1}^2
) \cO_{-1}
\Bigr\} \vec\phi
\nonumber
\enn
to obtain a constraint on the ope.
This amounts to a series of row and column operations on the
matrix equation
(\ref{eq.meq}).
If we expand the opes of $\phi, \tilde\phi$ in $z$ as
\eq
\phi(z) \vec{01;ss'}
 = \sum_{n\ge 0} z^{n-y}f_n
\;,\;\;
\tilde\phi(z) \vec{01;ss'}
 = \sum_{n\ge 0} z^{n-y-3}\tilde{f}_n
\,,
\en
where $y=h_{\yyy} +  h_{01,ss'} - h_{11,ss'}$,
and use the relations which come from (\ref{eq.vot})
we obtain
\eq \tilde{f}_n=\a_n f_n+ \sum_{p\ge 1} \left \{
\a Q_{-p}+[{p(\a^2 -3)+2(\a^2 -2s-s')\over 6}+n]L_{-p}
\right \}f_{n-p}
\,,
\label{eq.finalr}
\en
where
\eq
\a_n= - {1\over \a^2 }n(n-s)(n+\a^2 -s-s')
\,.
\en
We would like to impose $f_0 = \vec{11;ss'}$, $\tilde f_0 = 0$ and
$\tilde f_n \sim 0, n>0$.
We can in fact impose $\tilde f_0 = 0, n<s$, and solve for
$f_n, n<s$, but
we find
that the coefficient of $f_n$ vanishes explicitly
at $n=s$.  Nonetheless, for $n=s$ the right hand side of (\ref{eq.finalr}) is
perfectly well-defined, and, as a descendent primary state, it is
equal to the singular
vector in the module generated from $f_0$.

The recurrence relation which we obtain is of the same form as that
in the Virasoro case.
We can solve for $\tilde f_n$ and arrive at the following formula
which is analogous to that derived by Benoit and St-Aubin,
\eq
\cO_{1,s}\vec{11;ss'}
\!= \!\!
\sum_{p_i:\sum_{i=1}^r p_i=s}
\!\!
{{\prod_i
\left \{
\a Q_{-p_i}
+ [ \left( p_i(\a^2 +
3)+2(\a^2 -2s-s') \right)/6 + n_i ] L_{-p_i}
\right \} }\over
{\prod_{i=1}^{r-1} n_i(n_i - s )(n_i+\a^2-s-s')\a^{-2}} }
\vec{11;ss'}
\,,
\label{eq.final}
\en
where $n_i=\sum_{j=1}^{i} p_j$ and the product in the numerator is
ordered from right to left with increasing $i$.
Note that this formula is valid for arbitrary $s'$, not just
$s'\in\Bbb N$.

It is also possible to derive an alternative expression for this null vector
by swapping the order of the fields in the ope and using (\ref{eq.walt}) rather
than (\ref{eq.w}). This can be written
\eq
\cO_{1,s}\vec{11;ss'}
= \sum_{
{
\{ p_1,\ldots,p_r\}, p_i=1,2,3
}\atop{
\sum p_i=s
}}
{{\prod_i \a^2
\left \{
\Gamma^{n_i}_{p_i}
\right \} }\over
{\prod_i n_i(n_i-s)(n_i+\a^2-s-s')}}
\vec{11;ss'}
\,,
\label{eq.final2}
\en
where $n_i$ is as above, as before the numerator is ordered from right to left,
$p_i=1,2,3$ and
\eqq
\Gamma^{n}_1 &=& \a Q_{-1}+\big \{ -{3\over \a^2}(n^2+n[{\a^2\over 3}
-1-{2\over 3}(2s+s')])
\cont
+{(\a^2+1)\over 2}+{1\over 3}(s-s')-{1\over \a^2}[s+s'+1][s+1]
\big \}L_{-1}
\,,
\nonumber
\\
\Gamma^n_{2} &=& 2\a Q_{-2} - {1\over \a^2}(3n
-2s-s'-3)L_{-1}^2
-({(2s+s'-2\a^2)\over 3}+1 - n)L_{-2}
\,,
\nonumber
\\
\Gamma^n_3
&=& \a Q_{-3} -{1\over \a^2}L_{-1}^3 +L_{-1}L_{-2}+ {(\a^2-3)\over 6}
L_{-3}
\,.
\enn
We defer a rigorous proof that these vectors are indeed
null to our forthcoming paper \cite{BWat3}, together with more discussion
on the fusion of general representations of the $W_3$ algebra and other
algebras.

The formulae for the related null vectors at level $s$ in
representations $(11;s's)$, $(ss';11)$, $(s's;11)$ are easily obtained from the
expressions (\ref{eq.final}, \ref{eq.final2}) by making the substitutions
$\a \to -\a$, $\a \to -1/\a$, $\a\to 1/\a$
respectively.

\section{Conclusions}

We have shown how to obtain two different explicit formulae for a
subset of null vectors in the representation theory of the Virasoro
algebra and $W_3$ algebra. It is fairly clear how to generalise these
results to the algebras $W_n$ for $n>3$ for e.g.~the
fusion $(21...1;11...1)\times (01...1;p1...1)\to (11...1;p1...1)$.

One can also try applying the same
techniques to opes involving fields corresponding to other degenerate
representations. For example, in order to derive a matrix equation of
the form (\ref{eq.meq}) for the ope of the field $(31;11)$ with some
other primary field, we need to go to level five before the null
vectors and basis functions $P(m,n)$ balance in the required fashion.
In \cite{BWat3}, we shall address more rigorously the possibility of
defining covariant opes, their uniqueness or otherwise, and the
resulting fusion rules.

Finally it might prove interesting to try to reproduce these formulae from the
explicit null vectors in the affine $su(3)$ representations, as the
Virasoro null vectors have been reproduced from affine $su(2)$ null
vectors by Ganchev and Petkova \cite{GPet1}.

PB and GMTW would like to thank A.~Kent for numerous useful discussions.
PB would like to thank DAMTP, and the Newton Institute, Cambridge for
their kind hospitality while this letter was being written.

\end{document}